\documentstyle[preprint,aps]{revtex}
\begin{document}
\draft
\title{Minimal irreversible quantum mechanics. The decay of unstable states. }
\author{Diego G. Arb\'o,$^{1,2,3}$ Mario A. Castagnino,$^{1,2,4}$ Fabi\'an H. Gaioli,%
$^{1,5}$ Sergio Iguri$^1$ \\
$^1$Instituto de Astronom\'\i a y F\'\i sica del Espacio, C. C. 67, Suc. 28,%
\\
1428 Buenos Aires, Argentina\\
$^2$Consejo Nacional de Investigaciones Cient\'\i ficas y T\'ecnicas, Av.\\
Rivadavia 1917, 1033 Buenos Aires, Argentina\\
$^3$Departamento de Ingenier\'\i a e Investigaciones Tecnol\'ogicas,\\
Universidad Nacional de La Matanza, F. Varela 1903, 1754 S. Justo, Buenos\\
Aires, Argentina\\
$^4$Departamento de F\'\i sica, Facultad de Ciencias Exactas y Naturales,\\
Universidad de Buenos Aires, 1428 Buenos Aires, Argentina\\
$^5$Departamento de F\'\i sica, Universidad Nacional del Sur, Alem 1253,\\
8000 Bah\'\i a Blanca, Buenos Aires, Argentina}
\maketitle

\begin{abstract}
Brownian motion is modelled by a harmonic oscillator (Brownian particle)
interacting with a continuous set of uncoupled harmonic oscillators. The
interaction is linear in the coordinates and the momenta. The model has an
analytical solution that is used to study the time evolution of the reduced
density operator. It is derived in a closed form, in the one-particle sector
of the model. The irreversible behavior of the Brownian particle is
described by a reduced density matrix.
\end{abstract}

\section{Introduction}

\smallskip 

In previous works \cite{ijtp1,ijtp2,pa1,pa2} we have studied the time
evolution of a quantum oscillator coupled to a dense, but discrete (finite)
bath of harmonic oscillators. For such system an irreversible behavior has
appeared as a consequence of averaging in time the evolution of the
characteristic quantum-oscillator variables (macroscopic quantities), since
time evolution splits in very different scales: One related to small
fluctuations, which are erased by averaging, another one related with
recurrence phenomena, which are far enough of laboratory observational
times, and the last one connected with observable phenomena, which involves
irreversibility.

In Ref. \cite{ijtp1} it has been shown how to pass to the continuous bath. A
resonance, which can be isolated and leads to an evolution for a macroscopic
period of time (the same period as in the discrete case), has arisen because
of the energy of the quantum oscillator is embedded in the continuum. One
particularity of this continuous limit is that two of the three time scales
have become irrelevant since they have zero measure with respect to the
remaining time scale. However, this last scale is not exclusively governed
by the resonance (associated with an exponential decay) but also by
contributions coming from the semibounded feature of the energy spectrum.

Leaving these contributions aside the main behavior of the quantum
oscillator is an exponential decay towards the equilibrium with the bath,
and can be described only with the contribution due to the resonance. The
present work is the conclusion of the previous papers \cite
{LauraA,LauraB,LauraC,Castagnino}, where a novel method was used to work
directly in the continuum, including the exponential decay law in quantum
mechanics.\ In this paper we continue with the development of the formalism
we have called ``Minimal irreversible quantum mechanics,'' where time
asymmetry can be represented through the choice of a subspace of
``admissible'' or ``regular'' solutions of the evolution equation.

The main idea goes as follows. Usually rigorous quantum mechanics must be
formulated in a Gel'fand triplet \cite{Bogolubov} 
\begin{equation}
{\cal S\subset H\subset S^{\times }},  \label{3.1.1}
\end{equation}
where:

$\bullet {\cal S}$ is the space of ``regular states'' or test-functions
space, corresponding to Schwarz-class wave functions, that are considered as
the ``physical'' states.

$\bullet {\cal H}$ is the space of ``states,'' or Hilbert space, introduced
to extend the notion of probability to a larger space and to use the
well-known spectral theory of Hilbert spaces. These states correspond to
square-integrable wave functions.

$\bullet {\cal S^{\times }\ }$is the space of ``generalized states'', or
rigged Hilbert space, namely the space of linear (or antilinear) functionals
over ${\cal S}$, which are essentially used to find the spectral expansion
of the regular states (e.g. Fourier expansions).

\smallskip\ 

Let $K$ be the Wigner or time-inversion operator. As usual the evolution
Hamiltonian $H$ is time symmetric, i.e. 
\begin{equation}
KHK^{\dagger }=H.  \label{3.1.3}
\end{equation}

In the wave function representation the action of $K$ coincides with the
complex conjugation, so it is defined over ${\cal S}$ by 
\begin{equation}
K\varphi (x)=\varphi ^{*}(x).  \label{3.1.4a}
\end{equation}
Thus 
\begin{equation}
K:{\cal S\rightarrow S}.  \label{3.1.4b}
\end{equation}
Therefore ${\cal S}$ is also time symmetric.

But the real universe and macroscopic objects have clearly time-asymmetric
evolutions. Therefore the task of this paper and the preceding ones is to
explain how this time asymmetry appears while the quantum mechanical laws of
the universe (embodied in ${\cal H}$) are time symmetric. The usual and
successful explanation is based on coarse-graining: Macroscopic objects have
a huge number of dynamical variables and we can measure and control only a
small number of them, the so-called relevant variables. If we neglect the
rest of the variables, the irrelevant ones, we obtain time-asymmetric
evolution equations. Nevertheless in paper \cite{LauraA} (according to the
line of thought pioneered in Refs. \cite{Bohm}, \cite{Sudarshan}, \cite
{Ant.Prig}) we have follow a different way. We have developed a sort of
minimal irreversible quantum description, which reproduces time asymmetry
from the basic microscopic level directly, where the key point is the
presence of resonances (and additional hypotheses we have extracted from
Ref. \cite{pa1}).

Obviously we want to obtain the standard results making minimal changes to
the well established and usual quantum mechanics. If we change Eqs. (\ref
{3.1.3}) or (\ref{3.1.4a}), it is almost sure to find experimental problems.
So the minimal modification is to change Eq. (\ref{3.1.4b}) defining a new
test-functions space $\Phi _{+}\subset {\cal S}$ such that 
\begin{equation}
K:\Phi _{+}\rightarrow \Phi _{-}\neq \Phi _{+}.  \label{3.1.5}
\end{equation}
In this way $K$ is not even defined over the space of regular states $\Phi
_{+}$ and a time-asymmetric evolution arises.

This can be done if we postulate, as we have done in Ref. \cite{LauraA},%
\footnote{%
This postulate has been motivated in cosmological-global considerations in
Refs. \cite{Goslar} and \cite{Chile}.} that all the ``regular'' or
``admissible'' states belong to a space ${\cal H}_{+}\sim \theta (H_{+}^2)$
and also to ${\cal S}.$ Then $\Phi _{+}\sim \theta (H_{+}^2\cap {\cal S)}$ $%
[ $the time inverted states belong to a space ${\cal H}_{-}\sim \theta
(H_{-}^2)$ and $\Phi _{-}\sim \theta (H_{-}^2\cap {\cal S}),$ respectively]$%
, $where $\theta $ is the Heaviside step function that gives the restriction
to the positive real energy axis and $H_{\pm }^2$ are the Hardy class
function spaces \cite{Bohm}.\footnote{%
As spaces ${\cal H}_{-}$ and ${\cal H}_{+}$ are isomorphic they are normally
called ${\cal H}$ \cite{Bohm.Ga}.}

An ``irreversible'' quantum theory based on a Gel'fand triplet 
\begin{equation}
\Phi _{\pm }\subset {\cal H}_{\pm }\subset \Phi _{\pm }^{\times },
\label{3.1.6}
\end{equation}
is feasible and it yields physical results, as the dominant experimental
decay of unstable states, if the test-function space $\Phi _{+}$ is so
chosen. This will be valid for systems where the existence of resonances
dominates the evolution for the relevant period of observational time. We
have shown that, what it is done in the quoted papers \cite{Bohm}, \cite
{Sudarshan}, and \cite{Ant.Prig} is essentially a minimal modification of
the ordinary reversible quantum theory. In fact, from now on we will
consider that:

$\bullet \Phi _{+}$ is the space of ``regular states'' or test-functions
space, that are considered as the ``admissible'' states.

$\bullet {\cal H}_{+}$ is the space of ``states,'' or Hilbert space. These
states are again particular square-integrable wave functions. But in paper 
\cite{LauraA} and in this work we consider that only $\Phi _{+}$ contains
the ``admissible'' states.

$\bullet \Phi _{+}^{\times }{\cal \ }$is the space of ``generalized
states,'' or rigged Hilbert space, namely the space of linear (or
antilinear) functionals over $\Phi _{+},$ which are essentially used to find
the spectral expansion of the ``regular states'' in Sec. III.

The spaces with subscripts ``--'' contain the time-inverted states of the
corresponding spaces with subscripts ``+''.

Friedrichs model \cite{Friedrichs} was studied using this approach. In this
work, we show that this idea can be used to take a slightly different point
of view in studying dissipation phenomena of quantum Brownian motion. This
more complex model will force us to generalize the definition of space $\Phi
_{+}$ although the roles played by the characters in the triplets $\Phi
_{\pm }\subset {\cal H}_{\pm }\subset \Phi _{\pm }^{\times }$ will remain
the same.

Brownian motion has been extensively studied in the literature (we will only
quote those papers particularly relevant to our line of work). E.g., in Ref. 
\cite{FEJ} it was shown that for a system composed by a finite number of
linear interacting oscillators a dissipative behavior can be found in the
limit of a dense system (continuous spectrum). But, in this work we are
concerned directly with dense systems with continuous spectrum. The presence
of this continuous spectrum allows us to study the decay processes using
analytical properties familiar in scattering theory \cite{LauraB,JMY}. The
model is a very well-known and widely used system, consisting of a harmonic
oscillator coupled to an infinite and continuous bath. In this paper, as in
Refs. \cite{FEJ,HPZ}, the bath is composed by an infinite collection of
harmonic oscillators and the interaction is modelled to be linear and
characterized by the spectral weight, but otherwise arbitrary. We show that
the oscillator reaches a final equilibrium state via a damped evolution
which is mostly exponential. We also show that some deviations from this
exponential decay law (for very short and very long times) appear, which are
intimately related with the presence of a lower bound of the energy.

In Sec. II the whole system (single oscillator plus the bath) is described
and the Hamiltonian is introduced.

In Sec. III we diagonalize (in normal modes) the Hamiltonian. In the process
of diagonalization some problems emerge, such as the lost of the discrete
part of the energy spectrum \cite{Friedrichs,Arai}. We can bypass these
problems, if we use our definition of ``regular'' states. Then we $\ $can
perform an analytical continuation of the spectral decomposition of the
Hamiltonian, promoting the energy to complex values. To reach a successful
interpretation of the results we require to generalize the definition given
in paper \cite{LauraA} to the model we are now studying. The mathematical
bases of this generalization are shown in Appendix B.\footnote{%
The reader who is not familiar with rigged Hilbert spaces and functional
analysis can see Refs. \cite{Bohm} and \cite{Bogolubov}.} This appendix also
contains a rigorous mathematical understanding of the problem.

In Sec. IV mixed states and their evolution law are considered.

In Sec. V we deal with a very particular initial condition: An oscillator in
a zero-temperature bath. We find the reduced density operator and show that
the equilibrium state is reached. We accurately describe the time evolution
of the system and estimate the Zeno \cite{MS} and Khalfin \cite{Khalfin}
effects for very short and long times, respectively. These are the
deviations from an exact exponential decay law. Finally we show that our
solution satisfy a Lindblad master equation when discarding these deviations
from the exponential behavior.

Finally in Sec. VI some questions concerning irreversibility, already
considered in papers \cite{Gordo} and \cite{LauraA}, are discussed.

We state our conclusions in Sec. VII.

Three mathematical appendices complete this work.

\medskip\ \ 

\section{Particle-bath model}

\smallskip\ 

The system is a Brownian particle represented by a harmonic oscillator with
natural frequency $\Omega .$ It is well known that for a finite bath it is
not possible to prove convergence in an equilibrium state in the limit $%
t\rightarrow \infty $ because of the existence of recurrences \cite
{FEJ,Mazur,Rubin,Gruver}. However, for large systems these recurrence times
become extremely huge and we can eliminate them by passing to the limit of
an infinite continuous bath. Therefore, in this paper, we consider the
oscillator in contact with a bath, already modeled by a continuous set of
harmonic oscillators with natural frequencies $\omega .$ The coupling
between the system and the bath is assumed to be linear with strength $%
g(\omega )$. The Hamiltonian for the composite system, in terms of creation
and annihilation operators, is

\begin{equation}
H=\Omega a^{\dagger }a+\int_0^\infty d\omega \omega b_\omega ^{\dagger
}b_\omega +\lambda \int_0^\infty d\omega g(\omega )\left( a^{\dagger
}b_\omega +b_\omega ^{\dagger }a\right) .  \label{Hami}
\end{equation}
The first term corresponds to the system, the second to the bath, and the
third one corresponds to the interaction between them. In order that the
Hamiltonian would be positive definite we must require \cite{Gordo,Ullersma}
that $g(0)=0$ and

\begin{equation}
\Omega >\lambda ^2\int_0^\infty d\omega \frac{g^2(\omega )}\omega .
\label{posit}
\end{equation}
This is an important condition which selects the kind of spectral densities
appropriated to lead to an irreversible evolution. For example, the ohmic
case which is frequently used in the literature must be disregarded, unless
a cutoff is used. (Operators $b_\omega $ and $b_\omega ^{\dagger }$ are
rigorously defined in Appendix A).

The Fock basis is the tensor product of the Fock basis of the isolated
harmonic oscillator and those of the bath, namely 
\begin{equation}
\left| n,\omega _1...\omega _m\right\rangle =\left| n\right\rangle \otimes
\left| \omega _1...\omega _m\right\rangle ,  \label{base}
\end{equation}
where $\left| \omega _1...\omega _m\right\rangle $ represents a state with $%
m $ quanta in the bath, each one with frequency $\omega _j$ $(j=1,...,m)$.

The total number of quanta is conserved allowing us to solve the problem by
sectors (block diagonalization). The one-particle sector is referred as
Friedrichs' model \cite{Friedrichs} and contains the relevant information
that we need to compute physical quantities (see Sec. V).

\medskip\ \ 

\section{Normal modes of the Hamiltonian and analytic continuation}

\smallskip\ 

The linearity in the coupling term of $H$ allows us to easily find a new set
of uncoupled harmonic oscillators (normal modes), such that 
\begin{equation}
I=\int_0^\infty d\omega \ \widetilde{b}_\omega ^{\dagger }\widetilde{b}%
_\omega ,  \label{Hne-}
\end{equation}

\begin{equation}
H=\int_0^\infty d\omega \ \omega \widetilde{b}_\omega ^{\dagger }\widetilde{b%
}_\omega ,  \label{Hne}
\end{equation}
where

\begin{equation}
\widetilde{b}_\omega =\xi _\omega a+\int_0^\infty d\omega ^{\prime }\Phi
_\omega (\omega ^{\prime })b_{\omega ^{\prime }}.  \label{cuadrado}
\end{equation}
From a straightforward calculation \cite{LauraB,Sudarshan,Gordo}, using the
Heisenberg equations of motion, we obtain the coefficients of the unitary
change of variables which diagonalize the Hamiltonian, precisely

\begin{equation}
\Phi _\omega (\omega ^{\prime })=\delta (\omega -\omega ^{\prime })+\frac{%
\lambda \xi _\omega g(\omega ^{\prime })}{(\omega -\omega ^{\prime
}+i\varepsilon )}  \label{fid}
\end{equation}
and

\begin{equation}
\xi _\omega =\frac{\lambda g(\omega )}{\alpha (\omega +i\varepsilon )},
\label{psi}
\end{equation}
where 
\begin{equation}
\alpha (z)=z-\Omega -\lambda ^2\int_0^\infty d\omega \frac{g^2(\omega )}{%
z-\omega }.  \label{alfa}
\end{equation}
This function, which is the inverse of the reduced resolvent of $H$ in the
one-particle sector, is not entire because it has a cut along the positive
real axis corresponding to the continuous spectrum of the Hamiltonian. If $%
\alpha (z)\neq 0$ for all $z\in {\bf C}$, except for a possible real and
negative $\omega _0$ such that $\alpha (\omega _0)=0,$ an isolated solution
appears, which is non-analytic in $\lambda $ . We do not consider this case
henceforth, since we are interested in analytic solutions satisfying
condition (\ref{posit}).

If $\alpha (z)=0$ has no real solution it is not possible to find an
operator $\tilde a,$ such that $\tilde a\rightarrow a$ for $\lambda
\rightarrow 0.$ In this case we have lost the particle number operator
corresponding to the discrete part of the spectrum of $H$ and we do not have
the correct form of $H$ when $\lambda \rightarrow 0$ \cite{trio}. This
problem can be solved promoting the energy (or frequency) $\omega $ to be a
complex variable $z$. We define $\beta (z)\equiv \left[ \alpha (z)\right]
^{-1}$. It can be proved that $\beta (z)$ has the same analytic structure
than the one of the coefficient $S(z)$ of the scattering matrix \cite{Bohm}. 
$\beta (z)$ is a meromorphic function on a double Riemann sheet with a cut
along $[0,+\infty )$. $\beta _{\pm }(\omega )=\left[ \alpha _{\pm }(\omega
)\right] ^{-1}\equiv \beta (\omega \pm i\varepsilon )$ are defined on the
upper and lower half-planes of the first Riemann sheet $R_I$ (physical
sheet), and have meromorphic continuations to the lower and upper
half-planes, respectively, in the second sheet $R_{II}$ (unphysical sheet).
For simplicity we consider $g(z)$ such that the analytic extension of $\beta
_{+}(z)$ into the second sheet has a simple pole $z_0=\omega _0-\frac
i2\gamma $ [$\gamma >0$ and $\alpha _{+}(z_0)=0$] in the lower half-plane.
Also $\beta _{-}(z)$ has a simple pole $z_0^{*}$ on the upper plane in $%
R_{II}.$

We can now study the meaning of $z_0$. From the role played by $z_0$ in the
evolution equation we know that $({\rm Im}z_0)^{-1}=\gamma ^{-1}$ is the
mean life time of the unstable state $\left| 1,v\right\rangle =a^{\dagger
}\left| 0,v\right\rangle $ and $({\rm Re}z_0)$ is the shift of the bare
frequency $\Omega $ [see \cite{LauraA} and also Eq. (\ref{30})]. But, $z_0$
is the root of $\alpha _{+}(z)$ and from Eq. (\ref{alfa}) we can estimate,
up to the second order in $\lambda ,$ 
\begin{equation}
z_0=\Omega +\lambda ^2{\rm P}\int\limits_0^\infty d\omega \frac{g^2(\omega )%
}{\Omega -\omega }-i\pi \lambda ^2g^2(\Omega ),  \label{zetacero}
\end{equation}
where ${\rm P}$ denotes the Cauchy principal part of the integral.\footnote{%
The principal part comes from the well known identity between distributions
\par
\[
\frac 1{x+i\varepsilon }={\rm P}\frac 1x-i\pi \delta (x),\ x\in {\bf R} 
\]
} The mean life of the unstable state and the shift frequency are given by

\begin{equation}
\gamma =2\pi \lambda ^2g^2(\Omega )
\end{equation}
and 
\begin{equation}
\delta \Omega =\lambda ^2{\rm P}\int\limits_0^\infty d\omega \frac{%
g^2(\omega )}{\Omega -\omega }.  \label{shift}
\end{equation}
which are well-known expressions in the theory of unstable systems \cite
{Cohen}, usually derived from the Fermi golden rule.

Regarding the coupling function of the form $g(\omega )\sim \omega ^n$ we
find that the ohmic case ($n=1$) without cutoff does not satisfy the
positivity condition (\ref{posit}). If we call $\gamma _{1/2}$ the
coefficient for the subohmic case ($0<n<1$) and $\gamma _2$ the coefficient
for supraohmic case ($n>1$), it is easy to prove that 
\begin{equation}
\gamma _2\ll \gamma _1\ll \gamma _{1/2}.  \label{mm}
\end{equation}

Now we will find a generalized partition of the identity $I$ and a
generalized spectral decomposition of $H$ that recovers the discrete part of
the spectrum \cite{Sudarshan,Gordo,PP}. In order to do this let $\Gamma $ be
the curve of Fig. 1. It lays on $R_I$ for $\beta _{-}(z)$ and on $R_{II}$
for $\beta _{+}(z)$. We define the analytic function of $z\in {\bf C}$ 
\begin{equation}
\alpha _\Gamma (z)=z-\Omega -\lambda ^2\int_\Gamma dz^{\prime }\frac{%
g^2(z^{\prime })}{z-z^{\prime }},  \label{alfagama}
\end{equation}
which generalizes Eq. (\ref{alfa}).

To find the partition of the identity and a expansion of $H$ we will use in
some adequate analyticity properties.\footnote{%
Properties of this kind were already introduced in previous works \cite
{LauraA,Bohm,Ant.Prig,Gordo}.} Thus we will define a space $\Phi _{+}\subset 
{\cal H}_{+}$ of states $|\varphi \rangle $ such that the function $\langle
0,\omega _1,...,\omega _n|\varphi \rangle =\varphi _0(\omega _1,...,\omega
_n)$ would have an analytic continuation, for each variable $\omega _i$ ($%
1\leq i\leq n)$ to a region that include the singularity $z_0.$ This space $%
\Phi _{+}$ would be our space of ``regular,'' ``admissible'' or ``physical''
states. Precisely, generalizing what we have done in paper \cite{LauraA}, we
will chose $\Phi _{+}$ such that its states would satisfy Eq. (\ref{rea+})
of Appendix B. Analogously, the space $\Phi _{-}$ of the ``unphysical''
time-inverted states would satisfy Eq. (\ref{rea-}) of that appendix. With
this choice the analytic continuation that we will perform has a rigorous
meaning, since the operators act in a space $\Phi _{+}$which endowed with
adequate analytic properties. The demonstration of this fact is a
mathematical problem, which is considered in Appendices B and C, where we
generalize previous results. Then if $|\varphi \rangle \in \Phi _{+}$ and $%
|\psi \rangle \in \Phi _{-}$, from Eqs. (\ref{Hne-}) and (\ref{Hne}) using
appendix B and following the similar demonstration of paper \cite{LauraA},
it can be proved that 
\begin{eqnarray}
\langle \psi |\varphi \rangle &=&\langle \psi |(\widetilde{a}^{(-)\star }%
\widetilde{a}^{(-)}+\int_\Gamma dz\ \widetilde{b}_z^{(-)\star }\widetilde{b}%
_z^{(-)})|\varphi \rangle ,  \label{13'} \\
&&  \nonumber
\end{eqnarray}
\begin{equation}
\langle \psi |H|\varphi \rangle =\langle \psi |(z_0\widetilde{a}^{(-)\star }%
\widetilde{a}^{(-)}+\int_\Gamma dz\ z\widetilde{b}_z^{(-)\star }\widetilde{b}%
_z^{(-)})|\varphi \rangle .  \label{13''}
\end{equation}

The residue at $z_0$ contributes to the first terms of the r.h.s. of these
generalized partition of the identity and spectral decomposition of $H,$ as
in paper \cite{Sudarshan}, and, in a weak sense, the two previous equations
can be written as 
\begin{equation}
I=\widetilde{a}^{(-)\star }\widetilde{a}^{(-)}+\int_\Gamma dz\ \widetilde{b}%
_z^{(-)\star }\widetilde{b}_z^{(-)}  \label{13'''}
\end{equation}
\begin{equation}
H=z_0\widetilde{a}^{(-)\star }\widetilde{a}^{(-)}+\int_\Gamma dz\ z%
\widetilde{b}_z^{(-)\star }\widetilde{b}_z^{(-)},  \label{He}
\end{equation}
The creation and annihilation operators in all these equations reads

\begin{eqnarray}
\widetilde{a}^{(-)} &=&\frac 1{\sqrt{\alpha _{+}^{\prime }(z_0)}}\left[
a+\lambda \int_0^\infty dz\frac{g(z)}{[z_0-z]_{+}}b_z\right] ,  \nonumber \\
&&  \label{aaa} \\
\widetilde{a}^{(-)\star } &=&\frac 1{\sqrt{\alpha _{+}^{\prime }(z_0)}%
}\left[ a^{\dagger }+\lambda \int_0^\infty dz\frac{g(z)}{[z_0-z]_{+}}%
b_z^{\dagger }\right] ,  \nonumber  \label{ae}
\end{eqnarray}
and

\begin{eqnarray}
\widetilde{b}_\omega ^{(-)} &=&b_\omega +\frac{\lambda g(\omega )}{\eta
_{+}(\omega )}\left[ a+\lambda \int_0^\infty d\omega ^{\prime }\frac{%
g(\omega ^{\prime })}{\omega -\omega ^{\prime }+i\varepsilon }b_{\omega
^{\prime }}\right] ,  \nonumber \\
&&  \label{be} \\
\widetilde{b}_\omega ^{(-)\star } &=&b_\omega ^{\dagger }+\frac{\lambda
g(\omega )}{\alpha _{-}(\omega )}\left[ a^{\dagger }+\lambda \int_0^\infty
d\omega ^{\prime }\frac{g(\omega ^{\prime })}{\omega -\omega ^{\prime
}-i\varepsilon }b_{\omega ^{\prime }}^{\dagger }\right] .  \nonumber
\end{eqnarray}
The distribution $\frac 1{[z_0-z]_{+}}$ means

\begin{equation}
\int_0^\infty \frac{f(\omega )}{[z_0-\omega ]_{+}}d\omega =\int_\Gamma \frac{%
f(z)}{z_0-z}dz=\int_0^\infty \frac{f(\omega )}{z_0-\omega }d\omega +2\pi
if(z_0),  \label{a17}
\end{equation}
for every well-behaved analytical function $f(z).$ Observe that $\widetilde{b%
}_\omega ^{(-)\star }$ does not change if we replace $\int_\Gamma $ by $%
\int_0^\infty $ because, in this case, no pole is crossed, since $\alpha
_{-}(z)$ has not poles (in $S_I$). Nevertheless $\widetilde{b}_\omega ^{(-)}$
does change because the pole is crossed when we modify the integration
contour. We have shown this fact explicitly putting $\eta _{+}(\omega )$ in
place of $\alpha _{+}(\omega ),$ where 
\begin{equation}
\frac 1{\eta _{+}(\omega )}=\frac 1{\alpha _{+}(\omega )}+2\pi i\frac{\delta
(z-z_0)}{\alpha _{+}^{\prime }(z_0)},  \label{etamas}
\end{equation}
being $\delta (z-z_0)$ the extension of the Dirac delta defined {\it \`a la}
Gel'fand and Shilov \cite{Gel'Shi,CGGGAL}. As a consequence of these facts $%
\widetilde{a}^{(-)\star }\neq \widetilde{a}^{(-)\dagger }$ and $\widetilde{b}%
_\omega ^{(-)\star }\neq \widetilde{b}_\omega ^{(-)\dagger }$. The star
operation corresponds to the analytic generalization of the complex
conjugation, which, acting on an analytic function $f(z)$, is defined by%
\footnote{%
It corresponds to the symbol \# of paper \cite{LauraA}. Here we follow the
notation $\star $ of paper \cite{Gordo} in which we have also studied this
model.}

\begin{equation}
f^{\star }(z)=\left[ f(z^{*})\right] ^{*}.  \label{fzeta}
\end{equation}

From Eqs. (\ref{aaa}) and (\ref{be}) we see that we have four annihilation
operators due to the presence of complex eigenvalues of $H$ with the
corresponding doubling of solutions, since we have a pair of complex
conjugate values. They are generalized eigenvalues of the two analytic
continuations of $H$ into the lower ($-$) [upper ($+$)] -complex plane.
These operators are

\[
\widetilde{a}^{(-)};\ \widetilde{a}^{(+)};\ \widetilde{b}_\omega ^{(-)};\ 
\widetilde{b}_\omega ^{(+)}\label{aniq} 
\]
The vacuum is the state annihilated by any annihilation operator. The
Bogolubov transformation of Eqs. (\ref{aaa}) and (\ref{be}) does not mix
creation and annihilation operators, therefore the vacuum just defined is
actually the same state defined as the vacuum of the noninteracting
system+bath. So from Eq. (\ref{base}) the vacuum is the state $\left|
0\right\rangle \otimes \left| v\right\rangle \equiv \left| 0,v\right\rangle $%
, where $\left| v\right\rangle $ is the vacuum of the bath.

The corresponding creation operators are 
\[
\widetilde{a}^{(-)\star };\ \widetilde{a}^{(+)\star };\ \widetilde{b}_\omega
^{(-)\star };\ \widetilde{b}_\omega ^{(+)\star }\label{creac} 
\]

Starting from the common vacuum, by applying successively the operators $%
\widetilde{a}^{(-)\star }$ and $\widetilde{b}_\omega ^{(-)\star }$, the Fock
basis $\left\{ \left| \ \widetilde{}\ \right\rangle \right\} $ is built, and
with $\widetilde{a}^{(+)\star }$ and $\widetilde{b}_\omega ^{(+)\star }$ we
build up the Fock basis $\{|\ \overline{}$ $\rangle \}$. In the case of $%
\widetilde{a}^{(-)\star }$ and $\widetilde{a}^{(+)\star }$ the corresponding
vectors in the Fock bases of the one-particle sector are generalized
eigenvectors of $H$ with purely complex eigenvalues. They represent unstable
states, i.e. $\widetilde{a}^{(-)\star }|\widetilde{0,v}\rangle =\left| 
\widetilde{1,v}\right\rangle $ is a one-particle generalized eigenvector of $%
H$ corresponding to a complex eigenvalue $z_0$ and $\widetilde{a}^{(+)\star
}\left| \overline{0,v}\right\rangle =\left| \overline{1,v}\right\rangle $ is
a one-particle generalized eigenvector of $H$ corresponding to a complex
eigenvalue $z_0^{*}.$ In this way we are able to develop a second quantized
version of the theory of unstable states \cite{Gordo}.

Now we have two different number of quanta operators, $\widetilde{N}%
^{(-)}=\int_0^\infty d\omega \widetilde{b}_\omega ^{(-)\star }\widetilde{b}%
_\omega ^{(-)}$ and $\widetilde{N}^{(+)}=\int_0^\infty d\omega \widetilde{b}%
_\omega ^{(+)\star }\widetilde{b}_\omega ^{(+)},$ which are not Hermitian.
So two different Fock bases can be built satisfying

\begin{eqnarray}
\widetilde{N}^{(-)}\left| \widetilde{n,\omega _1...\omega _m}\right\rangle
&=&m\left| \widetilde{n,\omega _1...\omega _m}\right\rangle ,  \nonumber \\
&& \\
\widetilde{N}^{(+)}\left| \overline{n,\omega _1...\omega _m}\right\rangle
&=&m\left| \overline{n,\omega _1...\omega _m}\right\rangle .  \nonumber
\end{eqnarray}

The spectral decomposition of the Hamiltonian reads 
\begin{eqnarray}
H^{(-)} &=&\sum_{n=0}^\infty \sum_{m=0}^\infty \int_0^\infty
...\int_0^\infty d\omega _1...d\omega _m\left( z_0n+\omega _1+...+\omega
_m\right)  \nonumber \\
&&\ \hspace{1.0in}\times \left| \widetilde{n,\omega _1...\omega _m}%
\right\rangle \left\langle \overline{n,\omega _1...\omega _m}\right| ,
\label{descH}
\end{eqnarray}
which acts on the right of the Fock space generated by basis $\left\{ \left|
\ \widetilde{}\ \right\rangle \right\} $. But the ``same'' Hamiltonian can
also be written in the following way (using the other analytical
continuation, in which case it is evident that the next equation is only
weak, and it has a precise meaning operating between $|\varphi \rangle \in
\Phi _{-}$ and $|\psi \rangle \in \Phi _{+}$) 
\begin{eqnarray}
H^{(+)} &=&\sum_{n=0}^\infty \sum_{m=0}^\infty \int_0^\infty
...\int_0^\infty d\omega _1...d\omega _m\left( z_0^{*}n+\omega _1+...+\omega
_m\right)  \nonumber \\
&&\ \hspace{1.0in}\times \left| \overline{n,\omega _1...\omega _m}%
\right\rangle \left\langle \widetilde{n,\omega _1...\omega _m}\right| ,
\label{descHd}
\end{eqnarray}
which acts on the right of the Fock space generated by basis $\left\{ \left|
\ \overline{}\ \right\rangle \right\} $. In the same way, the identity reads 
\begin{eqnarray}
I^{(-)} &=&\sum_{n=0}^\infty \sum_{m=0}^\infty \int_0^\infty
...\int_0^\infty d\omega _1...d\omega _m\left| \widetilde{n,\omega
_1...\omega _m}\right\rangle \left\langle \overline{n,\omega _1...\omega _m}%
\right| ,  \nonumber  \label{I} \\
&&  \label{Id} \\
I^{(+)} &=&\sum_{n=0}^\infty \sum_{m=0}^\infty \int_0^\infty
...\int_0^\infty d\omega _1...d\omega _m\left| \overline{n,\omega
_1...\omega _m}\right\rangle \left\langle \widetilde{n,\omega _1...\omega _m}%
\right| .  \nonumber
\end{eqnarray}

As the eigenvalues of Eqs. (\ref{descH}) and (\ref{descHd}) are complex, in
order to deal with unstable states we must find an adequate mathematical
structure beyond the Hilbert space. In fact, these states are generalized
states. Thus, in the Appendix B we see that kets $\left| \ \widetilde{}\
\right\rangle $ and $\left| \ \overline{}\ \right\rangle $ are well defined
in a rigged Hilbert space formalism, i.e. they must be thought as antilinear
functionals acting on test spaces $\Phi _{\pm }$ and, as elements of a
vector space, they belong to the duals of $\Phi _{\pm },$ symbolized by $%
\Phi _{\pm }^{\times }.$ They define a double Gel'fand triplet structure $%
\Phi _{\pm }\subset {\cal H}_{\pm }\subset \Phi _{\pm }^{\times }$ \cite
{Bohm,Bohm.Ga,Gordo}.

What about energy conservation? The trouble emerges because the eigenvalues
of the Hamiltonian are now complex; thus some states decay in time (e.g.
vectors of $\Phi _{+}^{\times }$ which vanish for long times, see Sec. V).
Since some states vanish we may ask ourselves how the conservation of energy
can be possible. The answer is that energy is conserved anyhow. In order to
demonstrate this fact we will calculate the mean value of the Hamiltonian in
a state $\left| \varphi (t)\right\rangle
=\sum\limits_{nm}\int\limits_0^\infty ...\int\limits_0^\infty d\omega
_1...d\omega _mc_n(\omega _1...\omega _m)\left| \widetilde{n,\omega
_1...\omega _m}\right\rangle $, precisely 
\begin{eqnarray}
E &=&\left\langle \varphi (t)|H^{(-)}|\varphi (t)\right\rangle  \nonumber \\
\ &=&\sum\limits_{nn^{\prime }}\sum\limits_{mm^{\prime
}}\int\limits_0^\infty ...\int\limits_0^\infty d\omega _1...d\omega
_md\omega _1^{\prime }...d\omega _{m^{\prime }}^{\prime }c_n^{*}(\omega
_1...\omega _m)c_{n^{\prime }}(\omega _1^{\prime }...\omega _{m^{\prime
}}^{\prime })  \nonumber \\
&&\ \ \times e^{-i(z_0n^{\prime }+\omega _1^{\prime }+...+\omega _{m^{\prime
}}^{\prime })t}e^{i(z_0^{*}n+\omega _1+...+\omega _m)t}(z_0n+\omega
_1+...+\omega _m)  \nonumber \\
&&\ \ \times \left\langle \widetilde{n,\omega _1,...\omega _m}|\widetilde{%
n^{\prime },\omega _1^{\prime }...\omega _{m^{\prime }}^{\prime }}%
\right\rangle .  \label{ener}
\end{eqnarray}

Taking into account the orthogonality relations (\ref{conmu1}) and (\ref
{conmu4}) shown in Appendix B, and that $z_0=\omega _0-i\frac \gamma 2$, $%
\gamma >0$, Eq. (\ref{ener}) reduces to

\begin{equation}
E=\sum\limits_m\int\limits_0^\infty ...\int\limits_0^\infty d\omega
_1...d\omega _m\left| c_0(\omega _1...\omega _m)\right| ^2(\omega
_1+...+\omega _m),
\end{equation}
which is time independent. Thus energy is conserved. Conservation of the
norm and the number of particles can also be demonstrated in an analogous
way changing $H$ by $I$.

Finally we can observe that Eqs. (\ref{conmu1}) and (\ref{conmu4}) show that
the generalized eigenvectors have null norm and energy (with the exception
of those with $n=0$) \cite{LauraA,LauraC,Gordo,CGGL}. In the literature they
are called Gamow vectors, they are generalized states, and they represent
just idealized mathematical states (see Appendix B), as is the case of the
plane waves.

\medskip\ 

\section{Mixed states: Its evolution}

\smallskip\ 

A general pure state belonging to $\Phi _{+}$ (see Appendix B) can be
written as

\begin{equation}
\left| \Psi \right\rangle =\sum_n\sum_m\int_0^\infty ...\int_0^\infty
d\omega _1...d\omega _m\ c_n\left( \omega _1,...,\omega _m\right) \widetilde{%
\left| n,\omega _1...\omega _m\right\rangle },  \label{a}
\end{equation}
where $\widetilde{\left| n,\omega _1...\omega _n\right\rangle }\in \Phi
_{+}^{\times }$, and the most general density\footnote{%
More accurately, we would say, that it is the most general possible decaying
density matrix, as we will see.} matrix can be written as

\begin{eqnarray}
\rho &=&\sum_{nn^{\prime }}\sum_{mm^{\prime }}\int_0^\infty ...\int_0^\infty
d\omega _1...d\omega _md\omega _1^{\prime }...d\omega _{m^{\prime
}}^{\prime} c_{nn^{\prime }}\left( \omega _1,...,\omega _m,\omega
_1^{\prime} ,...,\omega _{m^{\prime }}^{\prime }\right)  \nonumber \\
&&\times \widetilde{\left| n,\omega _1...\omega _m\right\rangle }\widetilde{%
\left\langle n,\omega _1^{\prime },...,\omega _{m^{\prime }}^{\prime}
\right| }.  \label{dens}
\end{eqnarray}
If this $\rho $ is the initial state $\rho =\rho (0)$, the evolution law of $%
\rho (t)$ reads

\begin{equation}
\rho (t)=e^{-iH^{(-)}t}\rho (0)e^{iH^{(+)}t}.
\end{equation}
As $H$ is only self-adjoint in a generalized way \footnote{%
Recall that $H$ is self-adjoint in the Hilbert space, where ${\cal H=H}%
^{\times }$, but in the generalized Hilbert space this property essentially
becomes Eq. (\ref{footnote}).} \cite{Bohm} $H^{(-)}$ acts in a different way
than $H^{(+)}=H^{(-)\dagger }$ and there are right and left eigenvalues, 
\begin{eqnarray}
H^{(-)}\widetilde{\left| n,\omega _1...\omega _m\right\rangle }
&=&(z_0n+\omega _1+...+\omega _m)\widetilde{\left| n,\omega _1...\omega
_m\right\rangle },  \nonumber \\
&&  \label{footnote} \\
\left\langle \widetilde{n,\omega _1...\omega _m}\right| H^{(+)}
&=&(z_0^{*}n+\omega _1+...+\omega _m)\left\langle \widetilde{n,\omega
_1...\omega _m}\right| .  \nonumber
\end{eqnarray}
Then we have 
\begin{eqnarray}
\rho (t) &=&\sum\limits_{nn^{\prime }}e^{-\frac \gamma 2(n+n^{\prime
})t}e^{-i\omega _0(n-n^{\prime })t}\sum\limits_{mm^{\prime }}\int_0^\infty
...\int_0^\infty d\omega _1...d\omega _md\omega _1^{\prime }...d\omega
_{m^{\prime }}^{\prime }  \nonumber \\
&&\times e^{-i(\omega _1+...+\omega _m)t}e^{i(\omega _1^{\prime }+...+\omega
_{m^{\prime }}^{\prime })t}c_{nn^{\prime }}(\omega _1,...,\omega _m,\omega
_1^{\prime },...,\omega _{m^{\prime }}^{\prime })  \nonumber \\
&&\times \widetilde{\left| n,\omega _1...\omega _m\right\rangle }\widetilde{%
\left\langle n^{\prime },\omega _1^{\prime }...\omega _{m^{\prime }}^{\prime
}\right| }.  \label{30}
\end{eqnarray}

For an arbitrary initial state $\rho (0)$ a time dependent asymptotic ($%
t\rightarrow +\infty $) state is reached. The explanation of this fact is
simple. The modes of the bath are independent of each other [see Eq. (\ref
{Hami})], and so we cannot expect that the bath reaches equilibrium (cf.
Ref. \cite{LauraB}). Thus

\begin{eqnarray}
\rho (t) &\rightarrow &\rho _{*}(t)=\sum\limits_{mm^{\prime }}\int_0^\infty
...\int_0^\infty d\omega _1...d\omega _md\omega _1^{\prime }...
d\omega_{m^{\prime }}^{\prime }e^{-i(\omega _1+... +\omega
_m)t}e^{i(\omega_1^{\prime }+...+ \omega _{m^{\prime }}^{\prime })t} 
\nonumber \\
&&\times c_{00}(\omega _1,...,\omega _m,\omega _1^{\prime },...,
\omega_{m^{\prime }}^{\prime })\widetilde{\left| 0,\omega _1...
\omega_m\right\rangle }\left\langle \widetilde{0,\omega _1^{\prime }...
\omega_{m^{\prime }}^{\prime }}\right| .  \label{rhoinf}
\end{eqnarray}

For completeness we also write the evolution equation for the density
operator,

\begin{eqnarray}
\frac{d\rho (t)}{dt} &=&-i\sum\limits_{nn^{\prime }}e^{-\frac \gamma
2(n+n^{\prime })t}e^{-i\omega _0(n-n^{\prime })t}\sum\limits_{mm^{\prime
}}\int_0^\infty ...\int_0^\infty d\omega _1...d\omega _md\omega _1^{\prime
}...d\omega _{m^{\prime }}^{\prime }  \nonumber \\
&&\times e^{-i(\omega _1+...+\omega _m)t}e^{i(\omega _1^{\prime }+...+\omega
_{m^{\prime }}^{\prime })t}(z_0n+\omega _1+...+\omega _m-z_0^{*}n^{\prime
}-\omega _1^{\prime }-...-\omega _{m^{\prime }}^{\prime })  \nonumber \\
&&\times c_{nn^{\prime }}(\omega _1,...,\omega _m,\omega _1,...,\omega
_{m^{\prime }})\widetilde{\left| n,\omega _1...\omega _m\right\rangle }%
\left\langle \widetilde{n^{\prime },\omega _1^{\prime }...\omega _{m^{\prime
}}^{\prime }}\right| ,
\end{eqnarray}
which is clearly equal to 
\begin{equation}
\frac{d\rho (t)}{dt}=-i\left( H^{(-)}\rho -\rho H^{(+)}\right) =-iL\rho ,
\end{equation}
where $L$ is the generalized Liouvillian operator \cite{Gordo}. So we see
that the density operator follows an evolution described by a generalized
Liouville-von Neumann equation.

In spite of the result obtained in Eq. (\ref{rhoinf}) in Sec. V we show that
the reduced density operator $\rho _r,$ which is obtained by taking the
partial trace with respect to the environment modes, reaches equilibrium,
namely a time independent state. An equivalent way to find the equilibrium
state, closer to the spirit of our formalism, is to use a particular space
of observables, as in paper \cite{Decoherence}.

\medskip\ \ 

\section{Reduced density operator}

\smallskip\ 

As an illustration of the formalism we consider a simple example where the
initial state is a very particular state of the composed system,

\begin{equation}
\rho (0)=\rho _S(0)\otimes \rho _E(0),
\end{equation}
where

\begin{equation}
\rho _S(0)=c_{11}\left| 1\right\rangle \left\langle 1\right| +c_{10}\left|
1\right\rangle \left\langle 0\right| +c_{01}\left| 0\right\rangle
\left\langle 1\right| +c_{00}\left| 0\right\rangle \left\langle 0\right| 
\end{equation}
is the initial state of the discrete oscillator, or the initial reduced
density operator, (with $c_{11},c_{00}\geq 0$, $c_{11}+c_{00}=1$ and $%
c_{10}=c_{01}^{*}$) and

\begin{equation}
\rho _E(0)=\left| v\right\rangle \left\langle v\right|
\end{equation}
is the initial state of the bath, which does not have any quantum, namely,
it is in the ground state. This corresponds to a bath at zero temperature $%
T=0$ (in the zero and one-particle sector). The main features at any $T$ can
be reproduced but we begin with this example because the mathematical
computations are easier (recall that this model can be decomposed in sectors
of constant number of quanta). Also our initial conditions are such that
there is no correlation between the oscillator and the bath.

Our aim is to find the time dependence of the reduced matrix elements. It is
derived from the time evolution of the density operator

\begin{equation}
\rho (t)=e^{-iH^{(-)}t}\rho (0)e^{iH^{(+)}t}=e^{-iH^{(-)}t}I^{(-)}\rho
(0)I^{(+)}e^{iH^{(+)}t},
\end{equation}
where $I^{(-)}$ and $I^{(+)}$ are the identities in spaces $\Phi _{+}$ and $%
\Phi _{-}$ respectively\footnote{%
The difference in the conventions with respect to the use of $+$ and $-$ is
the following. In operators $+$ and $-$ are related with the analytic
continuations for $\pm i\varepsilon ,$ while $+$ and $-$ in spaces are
associated with the time evolution, which is only well defined for positive
or negative times, respectively.} [see Eqs. (\ref{Id})].

\begin{eqnarray}
\rho (t) &=&\left( e^{-iz_0t}\left| \widetilde{1,v}\right\rangle
\left\langle \overline{1,v}\right| +\int_0^\infty d\omega \ e^{-i\omega
t}\left| \widetilde{0,\omega }\right\rangle \left\langle \overline{0,\omega }%
\right| \right) \rho (0)  \nonumber \\
&&\times \left( e^{iz_0^{*}t}\left| \overline{1,v}\right\rangle \left\langle 
\widetilde{1,v}\right| +\int_0^\infty d\omega ^{\prime }\ e^{i\omega
^{\prime }t}\left| \overline{0,\omega ^{\prime }}\right\rangle \left\langle 
\widetilde{0,\omega ^{\prime }}\right| \right) .
\end{eqnarray}

We have only considered the terms of the identity that correspond to
zero-particle and one-particle subspaces, since, from the conservation of
the number of quanta, there is no contribution of other terms. We emphasize
that no approximations were carried out up to here.

Once we have the time evolution of the density operator, the following step
is to get the reduced density operator, tracing over the basis corresponding
to the environment,

\begin{eqnarray}
\rho _r(t) &=&tr_E\rho (t)=\sum\limits_{m=0}^\infty \int_0^\infty
...\int_0^\infty d\omega _1...d\omega _md\omega _1^{\prime }...d\omega
_m^{\prime }  \nonumber \\
&&\times \left\langle \omega _1...\omega _m\right| \rho (t)\left| \omega
_1^{\prime }...\omega _m^{\prime }\right\rangle \delta (\omega _1-\omega
_1^{\prime })...\delta (\omega _m-\omega _m^{\prime }),
\end{eqnarray}
where the $m=0$ term means $\left\langle v\right| \rho (t)\left|
v\right\rangle .$

As we have said, the contribution of terms $m=2,3,...$ vanishes. Therefore,
using the relations between ``new'' and ``old'' bases [Eqs. (\ref{aaa}) and (%
\ref{be})] and the conservation of trace, we obtain a positive definite
reduced density operator: 
\[
\rho _r(t)=c_{11}P(t)\left| 1\right\rangle \left\langle 1\right|
+c_{10}\Delta _0(t)\left| 1\right\rangle \left\langle 0\right| + 
\]

\begin{equation}
c_{01}\Delta _0^{*}(t)\left| 0\right\rangle \left\langle 1\right| +\left\{
c_{00}+c_{11}\left[ 1-P(t)\right] \right\} \left| 0\right\rangle
\left\langle 0\right| ,  \label{rhor}
\end{equation}
where

\begin{equation}
\Delta _0(t)=\frac{e^{-iz_0t}}{\alpha _{+}^{\prime }(z_0)}+\int_0^\infty
d\omega \ e^{-i\omega t}\frac{\lambda ^2g^2(\omega )}{\eta _{+}(\omega
)\alpha _{-}(\omega )},  \label{a(t)}
\end{equation}
and $P(t)=|\Delta _0(t)|^2$ is the survival probability of the state with
only one quantum in the discrete part.

We can write $P(t)$ as the sum of four terms where the first one, $\frac{%
e^{-\gamma t}}{|\alpha _{+}^{\prime }(z_0)|^2},$ shows an exact exponential
behavior. Expanding $|\alpha _{+}^{\prime }(z_0)|^{-2}$ as $1+O(\lambda ^2),$
we split the probability into two terms, one containing the purely
exponential contribution and the other that we call ``background,'' giving
rise to derivations from that purely exponential decay law, so that

\begin{equation}
P(t)=e^{-\gamma t}+{\rm background.}  \label{sinback}
\end{equation}
If we take a time neither very short nor very long, the background will be
smaller than the purely exponential term (for $\lambda \ll 1$) and it can be
neglected, which leads to an exponential decay-law. This is not true for
short times since $\frac{dP}{dt}(0)=0$, which leads to the so-called Zeno
effect \cite{Sudarshan}. For very long times the exponential term will decay
faster than the background will do, which is known as Khalfin effect \cite
{Khalfin}. We can force $P(t)$ to have an exponential appearance by defining
the decay rate $\Gamma (t)$ to be time dependent, namely

\begin{equation}
P(t)\equiv e^{-\Gamma (t)t},
\end{equation}
with

\[
\Gamma (t)=\gamma -\frac 1t\ln \left( 1+e^{\gamma t}\ {\rm background}%
\right) . 
\]
Obviously for an intermediate time the background can be neglected and $%
\Gamma (t)\simeq \gamma .$ The main restriction, imposed by the Zeno period,
is

\begin{equation}
\frac{dP}{dt}(0)=-\Gamma (0)=0.
\end{equation}
For a very long time the decay probability has also a non-exponential
contribution as a consequence of the semi-finiteness of the energy spectrum.
From papers \cite{LauraB,LauraC} we know that the survival amplitude goes to
zero as $t$ goes to infinity as a consequence of the Riemann-Lebesgue
theorem. Then the behavior of $\Delta _0$ depends on the small-frequency
behavior of $g^2(\omega ).$ For small $\omega ,$ $\alpha _{+}(\omega )\sim
-\omega _0,$ where condition (\ref{posit}) was also used. Then the form of $%
\Delta _0$ depends essentially on $g(\omega )$ for large $t.$ As an example
we consider the case where $g^2(\omega )\sim \omega ^n\exp \left( -\frac{%
\omega ^2}{\Lambda ^2}\right) ,$ where $\Lambda $ is a cutoff [see paragraph
before Eq. (\ref{mm})]. By evaluating the survival amplitude we have

\[
\Delta _0(t)=\lambda ^2\int_0^\infty d\omega \frac{g^2(\omega )}{\left|
\alpha _{+}(\omega )\right| ^2}e^{-i\omega t}\sim \int_0^{1/t}d\omega \
\omega ^ne^{-i\omega t}\exp \left( -\frac{\omega ^2}{\Lambda ^2}\right) , 
\]
where the contribution of high-frequency terms is negligible. Performing the
change of variables $\omega t=x,$ we obtain

\begin{equation}
\Delta _0(t)\sim t^{-(n+1)}\int_0^1dx\ x^ne^{-ix}\left( 1-\frac{x^2}{\Lambda
^2t^2}+\frac{x^4}{\Lambda ^4t^4}+...\right) .  \label{khalef}
\end{equation}
We can see that the survival amplitude merges into an algebraic long-time
tail. The first relevant contribution behaves as $t^{-n-1}$. As a
consequence the decay rate for long times must behave as 
\begin{equation}
\Gamma (t)\propto \frac{\ln t}t.
\end{equation}
The behavior at short times and intermediate times coincides with those
obtained in Ref. \cite{HPZ}. In Fig. 2 we show the qualitative behavior of $%
P(t).$ Zeno's time, $t_Z,$ and Khalfin's time, $t_K,$ are not in scale in
the picture in order to show the three different contributions to the decay
probability.

%
%
%

Eq. (\ref{rhor}) is the exact solution to the proposed problem, without
taking any approximation. The first, second, and third terms will vanish for 
$t\rightarrow \infty $; in fact $P(t\rightarrow \infty )=0$ and the same
happens for $\Delta _0(t)$ (recall that $P=|\Delta _0|^2$). The first term
of (\ref{a(t)}) has the factor $e^{-\frac \gamma 2t}$ and the second one
will tend to zero because of the Riemann-Lebesgue theorem. The probability
of having the vacuum will grow in time. This means that all quanta in the
discrete spectrum, except the ground state, decay into the continuum. So we
find the equilibrium reduced density operator

\begin{equation}
\rho _{*}=(c_{00}+c_{11})\left| 0\right\rangle \left\langle 0\right| =tr\rho
\left| 0\right\rangle \left\langle 0\right| =\left| 0\right\rangle
\left\langle 0\right| .  \label{rhoequi}
\end{equation}
As expected, the equilibrium state is the vacuum, namely for $t\rightarrow
\infty $ there are no quanta in the discrete spectrum, because the initial
quantum has decayed into the bath (the discrete oscillator has spread its
energy over the infinite oscillators of the bath with a distribution
centered at the shifted frequency $\omega _0$) \cite{LauraA,FEJ,LauraB}.
This means that the discrete harmonic oscillator has thermalized at $T=0$. A
similar result was recently obtained in Ref. \cite{Efimov}.

In order to check the compatibility of the solution found in Eq. (\ref{rhor}%
) we first briefly sketch the main points of the derivation we have done. We
have obtained the exact solution of the Liouville equation. As a particular
case, we have considered an initial condition restricted to the zero- and
one-particle sectors and we have traced this solution over the environment
modes. In that case, the survival probability $P(t)$ in Eq. (\ref{rhor}) can
be approximated by an exponential behavior, when the background contribution
is neglected. Now the solution for $\rho _r(t)$ obtained through this
approximation can be derived from a master equation of a Lindblad's form,
where the Lindblad generator is proportional to the destruction operator $a,$
since in our case we are only considering a zero-temperature bath in the
case of a damping motion caused by friction \cite{joos}:

\[
\dot \rho _r(t)=-i\Omega _0\left[ a^{\dagger }a,\rho _r\right] +\frac \gamma
2\left( 2a\rho _ra^{\dagger }-a^{\dagger }a\rho _r-\rho _ra^{\dagger
}a\right) .
\]
The probability of finding $n$ quanta follows a Pauli master equation:

\begin{equation}
\frac \partial {\partial t}\left\langle n\right| \rho _r\left|
n\right\rangle =\gamma \left[ \left( n+1\right) \left\langle n+1\right| \rho
_r\left| n+1\right\rangle -n\left\langle n\right| \rho _r\left|
n\right\rangle \right] .  \label{pauli}
\end{equation}
It is easy to see that $\rho _r$ of Eq. (\ref{rhor}), when the background is
neglected, is solution of the Pauli equation (\ref{pauli}).\footnote{%
Moreover $\rho _r(t)$ is possitive definites since this condition is
equivalent to $\rho _{11}\rho _{00}\geq |\rho _{10}|^2$ ($\rho
_{ij}=\left\langle i\right| \rho _r\left| j\right\rangle $) which is
obviously satisfied by $\rho _r(t)$ provided it is satisfied by $\rho _r(0)$.%
}

The results listed above are well known, but shown that our formalism works
as the usual well established theories on the subject.

\medskip\ 

\section{Semigroups, Wigner time-inversion and irreversibility}

\smallskip\ 

One of the features of the system we are studying is its irreversible
evolution; the system reaches the equilibrium at the far future, and of
course the inverse evolution is not possible anymore. These properties were
already found in Refs. \cite{LauraA,Gordo} and here they are briefly
reviewed.

The presence of a time-asymmetric behavior can be shown in two different
ways: As the splitting of the usual evolution group in two semigroups or as
the impossibility to make a time inversion. We consider that the second one
is the most eloquent.

\subsection{Semigroups}

Physical states can be chosen to be in test space $\Phi _{+},$ as we have
done (or, with a simple and physically irrelevant change of convention in
space $\Phi _{-}$), while corresponding generalized eigenvectors are in its
dual $\Phi _{+}^{\times }$ (or $\Phi _{-}^{\times }$). The proof is simple.
From the Paley-Wiener theorem \cite{Bohm} the following lemma can be deduced:

If $f(\omega )\in H_{+}^2$, $e^{i\omega t}f(\omega )\in H_{+}^2$ only for
positive times. Similarly if $f(\omega )\in H_{-}^2$, $e^{i\omega t}f(\omega
)\in H_{-}^2$ only for negative times.

The asymmetry in Hardy spaces can be immediately seen from the Fourier
transform representation. From the Paley-Wiener theorem it is known that
Hardy class functions from above can be represented as

\[
f(\omega )=\frac 1{\sqrt{2 \pi }}\int_0^\infty ds\ e^{i\omega s}\widehat{f}%
(s), 
\]
where 
\[
\widehat{f}(s)=\frac 1{\sqrt{2\pi }}\int_{-\infty }^\infty d\omega \
e^{-i\omega s}f(\omega ). 
\]

The Fourier transform of Hardy class functions $H_{+}^2$ is in the space of
square integrable functions with support on the positive real axis; and as
the Fourier transform of $e^{i\omega t}f(\omega )$ is given by 
\[
{\cal F}\{e^{i\omega t}f(\omega )\}=\widehat{f}(s-t), 
\]
i.e. a function with support on $[0,\infty )$ is transformed into a function
with support on $[t,\infty )$. For a negative time this last function has no
longer support on $[0,\infty )$ and therefore $e^{i\omega t}f(\omega )$ does
not belong to $H_{+}^2$. Analogously it can be proved the same property for
the Hardy class $H_{-}^2$.

To simplify, we analyze the one-particle case; generalization to $n$%
-particle states is straightforward. Let $\phi (\omega )$ be a function in $%
\theta \left[ {\cal S}\cap H_{\pm }^2\right] $ such that, as a consequence
of previous lemma, $e^{i\omega t}\phi (\omega )$ will also be in the same
space for $t>0$ only. As $\phi (\omega )=\langle \omega |\phi \rangle $ then
we can write in Dirac notation

\begin{equation}
e^{i\omega t}\phi (\omega )=e^{i\omega t}\langle \omega |\phi \rangle
=\langle \omega |e^{-iH^{(-)}t}|\phi \rangle ,
\end{equation}
and taking into account Eq. (\ref{defesp}) we can state that if $\left| \phi
\right\rangle \in \Phi _{+}$ then $e^{-iH^{(-)}t}|\phi \rangle \in \Phi _{+}$
only for $\ t>0$. In the same way, if $\left| \phi \right\rangle \in \Phi
_{-}$ then $e^{-iH^{(+)}t}|\phi \rangle \in \Phi _{-}$ for $t<0$. Then if we
postulate that $\Phi _{+}$ is the space of physical states and that the
physical evolution brings physical states into physical states, it turns out
that this evolution only take place in the period $t>0,$ so irreversibility
naturally appears \footnote{%
Moreover, states in $\Phi _{+}$ are linear combination of generalized
vectors $\left| \widetilde{n,\omega _1...\omega _m}\right\rangle $, and
these vectors evolve as 
\begin{equation}
e^{-iH^{(-)}t}\left| \widetilde{n,\omega _1...\omega _m}\right\rangle
=e^{-i(\omega _0n+\omega _1+...+\omega _m)t}e^{-\frac \gamma 2nt}\left| 
\widetilde{n,\omega _1...\omega _m}\right\rangle .  \label{evo1}
\end{equation}
Therefore, with the exception of $n=0$, they decay towards increasing $t$.
Then we say that physical vectors in $\Phi _{+}$ decay towards positive time
(except the states belonging to $\Phi _{+}\cap \Phi _{-},$ as the vacuum
state $\left| 0\right\rangle \bigotimes \left| v\right\rangle $, which does
not decay). Analogously 
\begin{equation}
e^{-iH^{(+)}t}\left| \overline{n,\omega _1...\omega _m}\right\rangle
=e^{-i(\omega _0n+\omega _1+...+\omega _m)t}e^{\frac \gamma 2nt}\left| 
\overline{n,\omega _1...\omega _m}\right\rangle ,  \label{evo2}
\end{equation}
and we can say that the unphysical states in $\Phi _{-}$ decay towards
negative times.}.

\subsection{The Wigner time-inversion}

The Wigner time-inversion operator acts in a real representation as \cite
{Messiah} 
\begin{equation}
K\varphi (x)=\varphi ^{*}(x).
\end{equation}
[cf. Eq. (\ref{3.1.4a})]. Then, as the complex conjugate of the functions of 
$H_{+}^2$ are the functions of $H_{-}^2$ we have that \cite{Gordo} 
\begin{eqnarray}
K &:&\Phi _{+}\rightarrow \Phi _{-}\neq \Phi _{+},  \nonumber \\
&& \\
K &:&\Phi _{-}\rightarrow \Phi _{+}\neq \Phi _{-}.  \nonumber
\end{eqnarray}

Then the Wigner operator is not well defined either within $\Phi _{+}$ or
within $\Phi _{-}$, so those states in $\Phi _{+}$ or $\Phi _{-}$ are not,
in general, $t-$symmetric. Therefore if we consider that only the states of $%
\Phi _{+}$ are ``physical'' or ``admissible'' the Wigner time inversion
transforms these states in ``unphysical'' ones and therefore it turns out to
be impossible since unphysical states simply do not exist in nature. Then,
through this mathematical structure, irreversibility is incorporated in our
theory.

Nevertheless $\left( \Phi _{+}\cap \Phi _{-}\right) $ is not an empty set 
\cite{Bohm.Ga}, so the time-inversion operator will be well defined there 
\begin{equation}
K:(\Phi _{+}\cap \Phi _{-})\rightarrow (\Phi _{+}\cap \Phi _{-}),
\label{intersec}
\end{equation}
and these states will describe reversible processes, which will be $t-$%
symmetric.

All these things lead us to the postulate of the introduction: ``Physical
states are in $\Phi _{+}$ ( or $\Phi _{-}$).'' In fact, this postulate
provides a mathematical structure to deal with irreversible process, as it
was also shown in papers \cite{Gordo,LauraA,LauraB,LauraC,Castagnino}. The
choice between $\Phi _{+}$ and $\Phi _{-}$ is conventional and does not lead
to physical consequences, but once we choose one of these spaces the
distinction between past and future becomes substantial. Moreover, if we
take into account the global structure of the universe, this choice can be
motivated from the asymmetry of this structure. (\cite{Goslar}, \cite{Chile})

\medskip\ 

\section{Concluding remarks}

\smallskip\ 

We outline the main results of this work.

We have diagonalized the full Hamiltonian of our model and extended it in
such a way that the solution is analytic in the interaction parameter. A
rigorous mathematical formalism can be introduced in order to deal with
unstable quantum systems (see the appendices).

Using this formalism, we have obtained a second quantized version of the
decay of unstable systems and we have found the corresponding creation and
annihilation operators of unstable states.

By means of a simple example, the exact time evolution of the reduced
density matrix at zero temperature has been studied. We have obtained an
exponential decay approach of $P(t)$ to the asymptotic value $P(t\rightarrow
\infty )=0$ which is expected when the particle is in thermal equilibrium
with a zero temperature bath. For short times we have found a quadratic
behavior for the decay probability $P(t)$ (Zeno effect). This short time
deviation from the exponential decay law was recently measured by first time
(see Ref. \cite{Wilkinson}). Other deviations from the exponential decay
law, in this case for long times, naturally arise in our framework: Khalfin
effect, which unfortunately are in practice far enough of any observational
time scale.

\smallskip 

\section{Acknowledgments}

We acknowledge the very helpful discussions with Edgardo Garc\'\i a Alvarez
and Roberto Laura. This work was partially supported by grants
CI1*-CT94-0004 and PSS$^{*}$ 0992 of European Union, PID 3183/93 of CONICET,
EX053 of Universidad de Buenos Aires, and 12217/1 of Fundaci\'on Antorchas.

\section{Appendices}

\appendix 

\section{Creation and annihilation operators}

Consider the annihilation and creation ``unsmeared operators,'' $b_\omega $
and $b_\omega ^{\dagger }$, respectively, that we have used in our
calculations. Commonly they are introduced in the mathematical framework of
quantum field theory by virtue of expressions like \cite{Bogolubov} 
\begin{equation}
b\left( \phi \right) =\int \phi \left( \omega \right) b_\omega \,d\omega
\label{one}
\end{equation}
or 
\begin{equation}
b^{\dagger }\left( \phi \right) =\int \phi ^{*}\left( \omega \right)
b_\omega ^{\dagger }\,d\omega ,  \label{two}
\end{equation}
where $b\left( \phi \right) $ and $b^{\dagger }\left( \phi \right) $ are the
(well-defined or ``smeared'') annihilation and creation operators of the
one-particle state $\phi \in {\cal H}$ (being ${\cal H}$ a Hilbert space)
and their action is interpreted as the annihilation, respectively the
creation, of a spectrum localized quanta, represented by a Dirac delta
centered in the real value $\omega $. In this context we find that 
\[
b^{\dagger }\left( \phi \right) :\bigoplus_{n=0}^\infty sym\left( {\cal H}%
^{\otimes n}\right) \rightarrow \bigoplus_{n=0}^\infty sym\left( {\cal H}%
^{\otimes n}\right) , 
\]
\begin{equation}
b^{\dagger }\left( \phi \right) \Phi =b^{\dagger }\left( \phi \right) \left(
\phi _0,\phi _1,\ldots ,\phi _n,\ldots \right) =\left( 0,b_0^{\dagger
}\left( \phi \right) \phi _0,b_1^{\dagger }\left( \phi \right) \phi
_1,\ldots ,b_n^{\dagger }\left( \phi \right) \phi _n,\ldots \right) ,
\end{equation}
where 
\begin{eqnarray}
b_n^{\dagger }\left( \phi \right) \phi _n &=&b_n^{\dagger }\left( \phi
\right) \left[ sym\left( \phi _n^{\left( 1\right) }\otimes \phi _n^{\left(
2\right) }\otimes \cdots \otimes \phi _n^{\left( n\right) }\right) \right] 
\nonumber \\
&=&sym\left( \phi _n^{\left( 1\right) }\otimes \phi _n^{\left( 2\right)
}\otimes \cdots \otimes \phi _n^{\left( n\right) }\otimes \phi \right)
\end{eqnarray}
and $b\left( \phi \right) =\left[ b^{\dagger }\left( \phi \right) \right]
^{\dagger }$.

Of course, in the framework of the Hilbert space foundation of quantum
mechanics ``definitions'' (\ref{one}) and (\ref{two}) are strictly formal.

It is evident that these equations are analogous to the formal definition of
the Dirac delta 
\begin{equation}
\phi \left( x\right) =\int \phi \left( \omega \right) \delta \left( x-\omega
\right) \,d\omega ,
\end{equation}
so, as in the case of the Dirac delta, the rigorous meaning of the
``unsmeared operators'' $b_\omega $ and $b_\omega ^{\dagger }$ must be found
in the rigged Hilbert space formulation of quantum mechanics.

One way to do this is considering the explicit definitions of $b_\omega $
and $b_\omega ^{\dagger }$ as distribution valued operators. Being ${\cal S}$
$\subset {\cal H}\subset {\cal S}^{\times }$ a rigged Hilbert space, we have
that 
\[
b_\omega :\bigoplus_{n=0}^\infty sym\left( {\cal S}^{\otimes n}\right)
\rightarrow \left[ \bigoplus_{n=0}^\infty sym\left( {\cal S}^{\otimes
n}\right) \right] ^{\times }, 
\]
\begin{equation}
b_\omega \Phi =b_\omega \left( \phi _0,\phi _1,\ldots ,\phi _n,\ldots
\right) =\left( b_{1\omega }\phi _1,b_{2\omega }\phi _2,\ldots ,b_{n\omega
}\phi _n,\ldots \right) ,
\end{equation}
where 
\begin{equation}
b_{n\omega }\phi _n=b_{n\omega }\left[ sym\left( \phi _n^{\left( 1\right)
}\otimes \phi _n^{\left( 2\right) }\otimes \cdots \otimes \phi _n^{\left(
n\right) }\right) \right]
\end{equation}
and $b_{n\omega }$ acts as in its definition given in Sec. II. Also 
\[
b_\omega ^{\dagger }:\bigoplus_{n=0}^\infty sym\left( {\cal S}^{\otimes
n}\right) \rightarrow \left[ \bigoplus_{n=0}^\infty sym\left( {\cal S}%
^{\otimes n}\right) \right] ^{\times }, 
\]
\begin{equation}
b_\omega ^{\dagger }\Phi =b_\omega ^{\dagger }\left( \phi _0,\phi _1,\ldots
,\phi _n,\ldots \right) =\left( 0,b_{0\omega }^{\dagger }\phi _0,b_{1\omega
}^{\dagger }\phi _1,\ldots ,b_{n\omega }^{\dagger }\phi _n,\ldots \right) ,
\end{equation}
where 
\begin{eqnarray}
b_{n\omega }^{\dagger }\phi _n &=&b_{n\omega }^{\dagger }\left[ sym\left(
\phi _n^{\left( 1\right) }\otimes \phi _n^{\left( 2\right) }\otimes \cdots
\otimes \phi _n^{\left( n\right) }\right) \right]  \nonumber \\
\ &=&sym\left( \phi _n^{\left( 1\right) }\otimes \phi _n^{\left( 2\right)
}\otimes \cdots \otimes \phi _n^{\left( n\right) }\otimes \delta _\omega
\right) .
\end{eqnarray}

Observe that, with these definitions, only the normal product is well
defined and that the ``$^{\dagger }$'' symbol is merely a convenient
notation, that is, there is not a rigorous Hermitian conjugation involved.
But if we want to define the canonical commutation relations we will be in
troubles, because the product of functionals is not uniquely defined. So
this is not the right way either.

Fortunately, there is another way to define the annihilation and creation
``operators,'' $b_\omega $ and $b_\omega ^{\dagger }$. This point of view is
more abstract than the previous one. Remember that the Dirac delta can be
considered as a tempered distribution, i.e. a continuous linear functional
on the one-particle regular states space ${\cal S}$. In an analogous way the
annihilation and creation ``operators,'' $b_\omega $ and $b_\omega
{}^{\dagger }$, and all the respective ``products'' can be considered as
continuous linear functionals on the Canonical Commutation Relations Algebra
= $CCR\left( {\cal S}\right) $.

We summarize some properties which characterize this kind of algebras \cite
{tesisig}.{\bf \ }Remember that, being ${\cal S}$ a nuclear metrizable
space, there exists a non-decreasing basis $\left\{ p_\alpha \right\}
_{\alpha \in I}$ of continuous seminorms such that each seminorm is
Hilbertian. Let us denote by ${\cal H}_\alpha $ the Hilbert space which is
the completion of the quotient space ${\cal S}/Ker\left( p_\alpha \right) $
with respect to the quotient norm $\widehat{p}_\alpha =p_\alpha /Ker\left(
p_\alpha \right) ,$ i.e. the space of equivalence classes defined by 
\[
\phi _\alpha =\left\{ \chi \in {\cal S}:p_\alpha \left( \phi -\chi \right)
=0\right\} , 
\]
where 
\[
\widehat{p}_\alpha :{\cal S}/Ker\left( p_\alpha \right) \rightarrow {\bf R}%
_{+}, 
\]
\[
\widehat{p}_\alpha \left( \phi _\alpha \right) =p_\alpha \left( \phi \right)
. 
\]
The $^{*}$-algebra $CCR\left( {\cal S}\right) $ is defined as the Hausdorff
projective limit \cite{Treves} of the collection $\left\{ CCR\left( {\cal H}%
_\alpha \right) \right\} _{\alpha \in I}$, where $CCR\left( {\cal H}_\alpha
\right) $ is the C$^{*}$-algebra \cite{harpe} generated by the family of
operators $\left\{ b\left( \phi \right) :\phi \in {\cal H}_\alpha \right\} ,$
with respect to the mappings that inject each $CCR\left( {\cal H}_\alpha
\right) $ into $CCR\left( {\cal H}_\beta \right) $ if $\alpha \geq \beta $,
where the order in $I$ is the induced by the ordering of the basis of
seminorms. We can characterize the $CCR\left( {\cal S}\right) $ as follows.
Since every projective limit of a collection of C$^{*}$-algebras is a b$^{*}$%
-algebra, i.e. a complete symmetric $^{*}$-algebra whose topology is defined
by a basis of continuous submultiplicative seminorms, $CCR\left( {\cal S}%
\right) $ is also a b$^{*}$-algebra (see Ref. \cite{igcas}). Moreover, we
have that $CCR\left( {\cal S}\right) $ is the strict inductive limit \cite
{Treves} of the collection $\left\{ \bigoplus_{j=0}^nsym\left( {\cal S}%
^{\otimes j}\right) \right\} _{n=0}^\infty $, so $CCR\left( {\cal S}\right) $
is a nuclear strict inductive limit of a collection of Fr\'echet spaces or $%
{\cal LF}^{*}$-algebra \cite{belanger}. With this we have that the algebra
is complete, barreled, and nuclear.

Finally we have that $CCR\left( {\cal S}\right) \subset CCR\left( {\cal H}%
\right) \subset \left[ CCR\left( {\cal S}\right) \right] ^{\times },$ which
represents a generalized Gel'fand triplet. So, viewing the relations (\ref
{one}) and (\ref{two}) as generalized expansions in the sense of the well
known Maurin's theorem, one can identify $b_\omega $ and $b_\omega ^{\dagger
}$ as continuous linear functionals on the algebra $CCR\left( {\cal S}%
\right) ,$ and we can say that they belong to $\left[ CCR\left( {\cal S}%
\right) \right] ^{\times }$.

\section{Rigged extension}

In this appendix we find a state space $\Phi $ with the required properties
to implement our formalism of unstable states. In order to adequately define
the vectors obtained in Sec. III we must restrict the Hilbert space, which
is the basic mathematical structure of ordinary quantum mechanics. Recall
that Dirac's bras are defined as linear functionals on kets of ${\cal H}.$%
\footnote{${\cal H}$ is the Hilbert space of the states we are considering.
It can be the whole space of states or some subspace with precise physical
properties, as the incoming or outgoing spaces.} These functionals belong to 
${\cal H}^{^{\times }},$ the topological dual of ${\cal H}.$ But in this
case ${\cal H}^{^{\times }}$ is isomorphic to ${\cal H},$ then one works
indistinguishably with kets and bras. However, if we restrict the topology
in order to take a dense subset $\Phi $ of ${\cal H},$ we break the
one-to-one correspondence between elements $\phi $ of $\Phi $ and continuous
linear or antilinear functionals $F$ over them. We will call $\Phi ^{\prime
} $ the dual space of linear functions and $\Phi ^{\times }$ the dual space
of antilinear functional. We usually use the latter one. It leads to a
triplet structure symbolized as $\Phi \subset {\cal H}\subset \Phi ^{\times
},$ where, to assure the convergence in the norm which defines the topology
of $\Phi ,$ we must require that $\left\langle \phi |F\right\rangle $ would
be finite \cite{Bohm,Bohm.Ga,Bogolubov}. This space $\Phi $ will be the
space of ``regular'' states $\Phi _{+},$ as we have explained above.
Changing the our convention it can be $\Phi _{-}$

In our case, a necessary condition for $\left| \phi \right\rangle \in \Phi $
is that, the following expression, a generalization of Eq. (\ref{aaa}),
would have a rigorous meaning,

\begin{eqnarray}
\left\langle \phi |\widetilde{n,v}\right\rangle &=&\left\langle \phi \left|
\left[ \widetilde{a}^{(-)\star }\right] ^n\right| 0,v\right\rangle  \nonumber
\\
&=&\frac 1{[\alpha ^{\prime }(z_0)]^{\frac n2}}\left[ \left\langle \phi
|nv\right\rangle +\lambda \int_0^\infty d\omega _1\frac{g(\omega _1)}{%
[z_0-\omega _1]_{+}}\left\langle \phi |n-1,\omega _1\right\rangle +\ldots
\right.  \label{fi.n} \\
&&\left. +\lambda ^n\int_0^\infty \ldots \int_0^\infty d\omega _1\ldots
d\omega _n\frac{g(\omega _1)}{[z_0-\omega _1]_{+}}\cdots \frac{g(\omega _n)}{%
[z_0-\omega _n]_{+}}\left\langle \phi |0\omega _1...\omega _n\right\rangle
\right] .  \nonumber
\end{eqnarray}

The last term of the second member of Eq. (\ref{fi.n}) must be well defined,
so the function $\left\langle \phi |0,\omega _1...\omega _n\right\rangle
=\phi _0^{*}(\omega _1...\omega _n)$ has to have an analytic continuation in
each variable $\omega _i$ ($0\leq i\leq n)$ to a region which includes the
singularity $z_0$, so that the integral defines an analytic $n$-dimensional
function evaluated at $z_0$.

The simplest choice for $\left\langle \phi |0,\omega _1...\omega
_n\right\rangle $ which does not depend on the localization of $z_0$ is that 
$\left\langle \phi |0,\omega _1...\omega _n\right\rangle $ would be a Hardy
function from below $H_{-}^2$ \cite{Bohm.Ga} for each variable. It is
equivalent to

\begin{equation}
\left\langle \phi |0,\omega _1...\omega _n\right\rangle \in \theta ({\cal S}%
\cap H_{-}^2)^{\otimes n}.  \label{cuida}
\end{equation}

This generalizes the criterium previously used for the one-particle sector 
\cite{LauraA,Gordo,Ant.Prig}. From this criterium it can be proved that all
the mathematical expressions above are well defined (see Appendix C). Then
in order to $\left\langle i,\omega _1...\omega _m|\phi \right\rangle $ $%
(i+m=n,$ $n\in {\cal N})$ be well defined, it must belong to the following
function space

\begin{equation}
\left\langle i,\omega _1...\omega _m|\phi \right\rangle \in
\bigoplus\limits_{m=0}^\infty \theta \left[ {\cal S}\cap H_{+}^2\right]
^{\otimes m},  \label{rea+}
\end{equation}
where ${\cal S}$ is the Schwartz space \cite{Bohm.Ga,Bogolubov}, and $\theta 
$ is the Heaviside step function, which gives the restriction to the
positive real axis.

If we do the same in order to define $\left\langle \phi |\overline{n,v}%
\right\rangle $, we find another realization space

\begin{equation}
\left\langle i,\omega _1...\omega _m|\phi \right\rangle \ \in
\bigoplus\limits_{m=0}^\infty \theta \left[ {\cal S}\cap H_{-}^2\right]
^{\otimes m},  \label{rea-}
\end{equation}
with $(i+m=n,$ $n\in {\cal N})$. Therefore we define the following spaces

\begin{equation}
\Phi _{\pm }=\left\{ \phi \ /\ \left\langle i,\omega _1...\omega _m|\phi
\right\rangle \in \bigoplus\limits_{m=0}^\infty \theta \left[ {\cal S}\cap
H_{\pm }^2\right] ^{\otimes m}\ \right\} .  \label{defesp}
\end{equation}

The generalized eigenvectors belong to the dual spaces $\Phi _{\pm }^{\times
}$, since they are antilinear functionals \cite{Bohm} on spaces (\ref{rea+})
and (\ref{rea-}), 
\begin{eqnarray}
\left| i,\widetilde{\omega _1...\omega _m}\right\rangle &\in &\Phi
_{+}^{\times },  \nonumber \\
&& \\
\left| \overline{i,\omega _1...\omega _m}\right\rangle &\in &\Phi
_{-}^{\times }.  \nonumber
\end{eqnarray}
These generalized eigenvectors fulfill the following relations (see \cite
{LauraA})

\begin{eqnarray}
\left\langle i,\widetilde{\omega _1...\omega _m}|\widetilde{i^{\prime
},\omega _1^{\prime }...\omega _{m^{\prime }}^{\prime }}\right\rangle &=&0, 
\nonumber \\
&&  \label{conmu1} \\
\left\langle \overline{i,\omega _1...\omega _m}|\overline{i^{\prime },\omega
_1^{\prime }...\omega _{m^{\prime }}^{\prime }}\right\rangle &=&0,  \nonumber
\label{conmu2}
\end{eqnarray}

\begin{equation}
\left\langle \overline{i,\omega _1...\omega _m}|\widetilde{i^{\prime
},\omega _1^{\prime }...\omega _{m^{\prime }}^{\prime }}\right\rangle =\frac{%
\delta _{ii^{\prime }}\delta _{mm^{\prime }}}{(m!)^2}\sum\limits_{\sigma \in
G_p}\sum\limits_{\tau \in G_p}\delta \left( \omega _{\sigma _1}^{\prime
}-\omega _{\tau _1}\right) ...\delta \left( \omega _{\sigma _m}^{\prime
}-\omega _{\tau _m}\right) .  \label{conmu3}
\end{equation}
Eqs. (\ref{conmu1}) are valid except for $i=i^{\prime }=0,$ and in this case
we have 
\begin{eqnarray}
\left\langle \widetilde{0,\omega _1...\omega _n}|\widetilde{0,\omega
_1^{\prime }...\omega _{n^{\prime }}^{\prime }}\right\rangle &=&\frac{\delta
_{nn^{\prime }}}{(n!)^2}\sum\limits_{\sigma ,\tau \in G_p}\delta \left(
\omega _{\sigma _1}^{\prime }-\omega _{\tau _1}\right) ...\delta \left(
\omega _{\sigma _n}^{\prime }-\omega _{\tau _n}\right) ,  \nonumber \\
&&  \label{conmu4} \\
\left\langle \overline{0,\omega _1...\omega _n}|\overline{0,\omega
_1^{\prime }...\omega _{n^{\prime }}^{\prime }}\right\rangle &=&\frac{\delta
_{nn^{\prime }}}{(n!)^2}\sum\limits_{\sigma ,\tau \in G_p}\delta \left(
\omega _{\sigma _1}^{\prime }-\omega _{\tau _1}\right) ...\delta \left(
\omega _{\sigma _n}^{\prime }-\omega _{\tau _n}\right) .  \nonumber
\end{eqnarray}
$G_p$ is the group of permutations. Eqs. (\ref{conmu1}) say that the norm of
generalized eigenvectors is zero (except $i=0$), which is a necessary fact
to conserve energy \cite{Gordo,LauraA}. It is not contradictory to have
null-norm vectors because these are generalized vectors which are not in the
usual Hilbert space and have an underlying indefinite metric structure \cite
{Gordo}. If we define the spaces ${\cal H}_{+}$ and ${\cal H}_{-}$ as the
spaces $\Phi _{-}$ and $\Phi _{+}$ where the condition about ${\cal S}$ is
not required, we have 
\[
{\cal H}_{\pm }=\left\{ \phi \ /\ \left\langle n,\omega _1...\omega
_m...|\phi \right\rangle \in \bigoplus\limits_{m=0}^\infty \theta \left[
H_{\pm }^2\right] ^{\otimes m}\ \right\} , 
\]
then we arrive to the triplets under Eq. (\ref{Id}) and in Eq. (\ref{RHS}).%
\footnote{%
In this way $\Phi _{\pm }$ is dense in ${\cal H}_{\pm }$ which is the
outgoing (incoming) space \cite{LyP}. The ${\cal H}$ cited in paper \cite
{LauraA} is actually the outgoing space ${\cal H}_{+}.$}

Using Eq. (\ref{defesp} ), we find a double structure of rigged Hilbert
spaces for our model,

\begin{eqnarray}
\Phi _{+} &\subset &{\cal H}_{+}\subset \Phi _{+}^{\times },  \nonumber \\
&&  \label{RHS} \\
\Phi _{-} &\subset &{\cal H}_{-}\subset \Phi _{-}^{\times }.  \nonumber
\end{eqnarray}

\section{The double integral theorem}

In Ref. \cite{Ant.Prig} it was demonstrated that if we want that the
integral 
\begin{equation}
\int_{{\bf R}^{+}}d\omega \frac{g(\omega )\langle \phi |\omega \rangle }{%
z_0-\omega }  \label{C.1}
\end{equation}
would be well defined, it is sufficient that 
\begin{equation}
\langle \phi |\omega \rangle \in \theta ({\cal S}\cap H_{-}^2).  \label{C.2}
\end{equation}
In the two variables case it is the integral 
\begin{equation}
\int_{{\bf R}^{+}}d\omega \frac{g(\omega )}{z_0-\omega }\int_{{\bf R}%
^{+}}d\omega ^{\prime }\frac{g(\omega ^{\prime })\langle \phi |\omega
,\omega ^{\prime }\rangle }{z_0-\omega }  \label{C.3}
\end{equation}
the one that must be well defined. In this case we prove the following
theorem.

{\bf Theorem.}

If 
\begin{equation}
\phi (\omega ,\omega ^{\prime })=\langle \phi |\omega ,\omega ^{\prime
}\rangle \in \theta ({\cal S}\cap H_{-}^2)^{\otimes 2},  \label{C.4}
\end{equation}
then integral (\ref{C.3}) is well defined.

{\bf Proof.}

If condition (\ref{C.4}) is fulfilled, as ${\cal S}$ is a Fr\'echet space%
\footnote{%
A Fr\'{e}chet space is a metrizable complete space.} we have (\cite{Treves},
page 459) 
\begin{equation}
\phi (\omega ,\omega ^{\prime })=\sum_{i=0}^\infty \lambda _i\phi
_1^i(\omega )\phi _2^i(\omega ^{\prime }),  \label{C.5}
\end{equation}
where $\sum_{i=0}^\infty |\lambda _i|<1,\phi _1^i(\omega ),$ $\phi
_2^i(\omega )\in \theta ({\cal S}\cap H_{-}^2)$ $(i=1,2,...)$, $\phi _1^i,$ $%
\phi _2^i\rightarrow 0$ when $i\rightarrow \infty ,$ and the r.h.s. of Eq. (%
\ref{C.5}) is absolutely convergent, namely the series 
\[
\sum_{i=0}^\infty p(\lambda _i\phi _1^i\phi _2^i) 
\]
is convergent for any continuous seminorm $p$ over $\theta ({\cal S}\cap
H_{-}^2)^{\otimes 2}.$ Let us now define the seminorm $p_{z_0}$ as 
\begin{equation}
p_{z_0}(\phi )=D^2\int_{{\bf R}^{+}}d\omega \frac{|g(\omega )|}{|z_0-\omega |%
}\int_{{\bf R}^{+}}d\omega ^{\prime }\frac{|g(\omega ^{\prime })||\langle
\phi |\omega ,\omega ^{\prime }\rangle |}{|z_0-\omega ^{\prime }|},
\label{C.6}
\end{equation}
where $D$ is the distance from ${\bf R}_{+}$ to $z_0$ ($D=\gamma /2$). We
must demonstrate that $p_{z_0}$ is a continuous seminorm. We use the
H\"older inequality \cite{Arbo3} 
\begin{equation}
\left\| \frac{g(\omega )}{z_0-\omega }\frac{g(\omega ^{\prime })\langle \phi
|\omega ,\omega ^{\prime }\rangle }{z_0-\omega }\right\|
_1=D^{-2}p_{z_0}(\phi )\leq \left\| g(\omega )g(\omega ^{\prime })\phi
(\omega ,\omega ^{\prime })\right\| _1\left\| \frac 1{(z_0-\omega
)(z_0-\omega ^{\prime })}\right\| _\infty  \label{C.7}
\end{equation}
for any $z_0\in {\bf C}_{-}$ (the lower half-plane).\footnote{%
We remember that 
\[
\left\| f\right\| _1=\int_{{\bf R}_{+}}|f(\omega )|d\omega , 
\]
\[
\left\| f\right\| _\infty =\sup \{|f(\omega )|;\omega \in {\bf R}_{+}\}, 
\]
and also that ${\cal S\in }L_1$ , i.e. any Schwartz function is integrable.}
Since 
\begin{equation}
\left\| \frac 1{(z_0-\omega )(z_0-\omega ^{\prime })}\right\| _\infty =\sup
\left\{ \left| \frac 1{(z_0-\omega )(z_0-\omega ^{\prime })}\right| ;\omega
,\omega ^{\prime }\in {\bf R}_{+}\right\} =D^{-2}  \label{C.8}
\end{equation}
Equation (\ref{C.7}) reads 
\begin{equation}
p_{z_0}(\phi )\leq \left\| g(\omega )g(\omega ^{\prime })\phi (\omega
,\omega ^{\prime })\right\| _1  \label{C.9}
\end{equation}
for any $z_0\in {\bf C}_{-}.$ Then $p_{z_0}(\phi )$ is not only a continuous
seminorm but also a continuous norm over $\theta ({\cal S}\cap
H_{-}^2)^{\otimes 2}.$ Then 
\[
\sum_{i=0}^\infty p_{z_0}[\lambda _i\phi _1^i(\omega )\phi _2^i(\omega
^{\prime })]=D^2\sum_{i=0}^\infty |\lambda _i|\int_{{\bf R}^{+}}d\omega 
\frac{|g(\omega )|}{|z_0-\omega |}\int_{{\bf R}^{+}}d\omega ^{\prime }\frac{%
|g(\omega ^{\prime })||\phi _1^i(\omega )||\phi _2^i(\omega ^{\prime })|}{%
|z_0-\omega ^{\prime }|} 
\]
\begin{equation}
=D^2\sum_{i=0}^\infty \int_{{\bf R}^{+}}d\omega \int_{{\bf R}^{+}}d\omega
^{\prime }\left| \lambda _i\frac{g(\omega )}{z_0-\omega }\frac{g(\omega
^{\prime })}{z_0-\omega ^{\prime }}\phi _1^i(\omega )\phi _2^i(\omega
^{\prime })\right| <\infty .  \label{C.10}
\end{equation}
So, calling 
\begin{equation}
f_i^{z_0}(\omega ,\omega ^{\prime })=\lambda _i\frac{g(\omega )}{z_0-\omega }%
\frac{g(\omega ^{\prime })}{z_0-\omega ^{\prime }}\phi _1^i(\omega )\phi
_2^i(\omega ^{\prime }),  \label{C.11}
\end{equation}
from the corollary of the Lebesgue theorem (\cite{Arbo18}, page 33) we know
that, if the series 
\[
\sum_{i=0}^\infty f_i^{z_0}(\omega ,\omega ^{\prime })<\infty 
\]
converge a.e. in ${\bf R}_{+}\times {\bf R}_{+}$, then the series,
considered as a function of $(\omega ,\omega ^{\prime }),$ belongs to $L_1$
and 
\begin{equation}
\int_{{\bf R}^{+}}d\omega \int_{{\bf R}^{+}}d\omega ^{\prime
}\sum_{i=0}^\infty f_i^{z_0}(\omega ,\omega ^{\prime })=\sum_{i=0}^\infty
\int_{{\bf R}^{+}}d\omega \int_{{\bf R}^{+}}d\omega ^{\prime
}f_i^{z_0}(\omega ,\omega ^{\prime }).  \label{C.12}
\end{equation}
Then, going back to Eq. (\ref{C.3}) we have 
\[
\int_{{\bf R}^{+}}d\omega \frac{g(\omega )}{z_0-\omega }\int_{{\bf R}%
^{+}}d\omega ^{\prime }\frac{g(\omega ^{\prime })\langle \phi |\omega
,\omega ^{\prime }\rangle }{z_0-\omega }=\int_{{\bf R}^{+}}d\omega \frac{%
g(\omega )}{z_0-\omega }\int_{{\bf R}^{+}}\sum_{i=0}^\infty \lambda _i\phi
_1^i(\omega )\phi _2^i(\omega ^{\prime }) 
\]
\begin{equation}
=\sum_{i=0}^\infty \lambda _i\int_{{\bf R}^{+}}d\omega \frac{g(\omega )\phi
_1^i(\omega )}{z_0-\omega }\int_{{\bf R}^{+}}d\omega ^{\prime }\frac{%
g(\omega ^{\prime })\phi _2^i(\omega ^{\prime })}{z_0-\omega }.  \label{C.14}
\end{equation}
The l.h.s. of Eq. (\ref{C.14}) is well defined since it is an integral of a $%
L_1$ function. Moreover it is the sum of products of two well defined
integrals, like the one of Eq. (\ref{C.1}), since from hypothesis $\phi
_1^i(\omega ),\phi _2^i(\omega ^{\prime })\in \theta ({\cal S}\cap H_{-}^2).$
Thus, the proof is complete.

Of course, this theorem can be generalized from the case of two factors, to
the case of $n$ factors and, taking into account Eq. (\ref{a17}), it proves
that Eq. (\ref{fi.n}) is well defined if condition (\ref{cuida}) is
fulfilled.

\section{Figure captions}

Fig. 1: Deformation of the contour of integration taking into account the
presence of the complex pole $z_0.$

\smallskip\ 

\noindent Fig. 2: Behavior of the decay probability showing the Zeno,
exponential, and Khalfin phases.

\medskip\

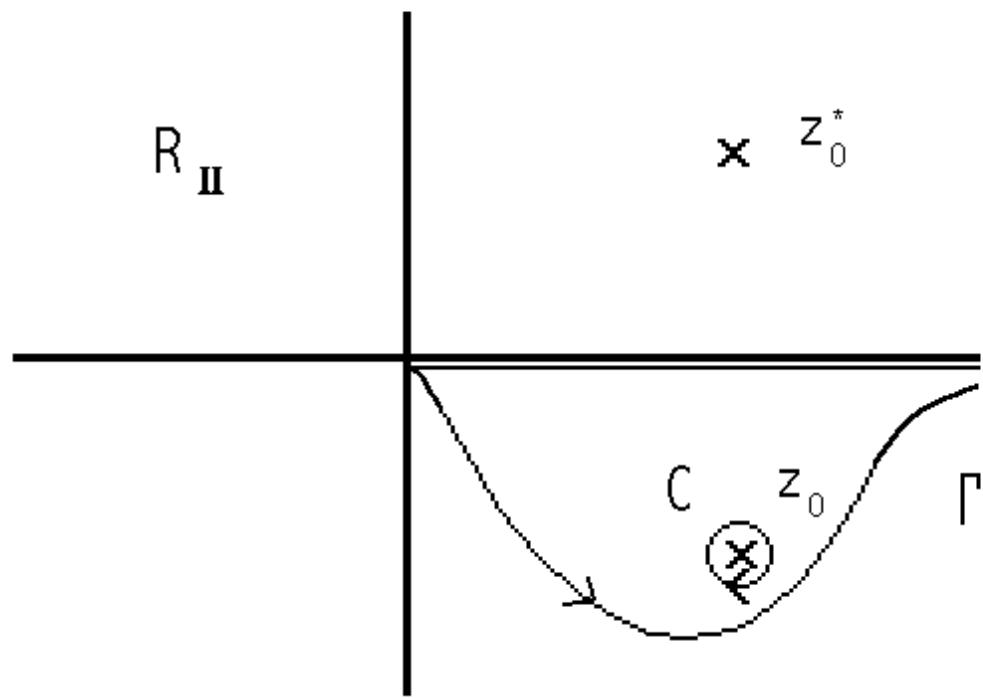
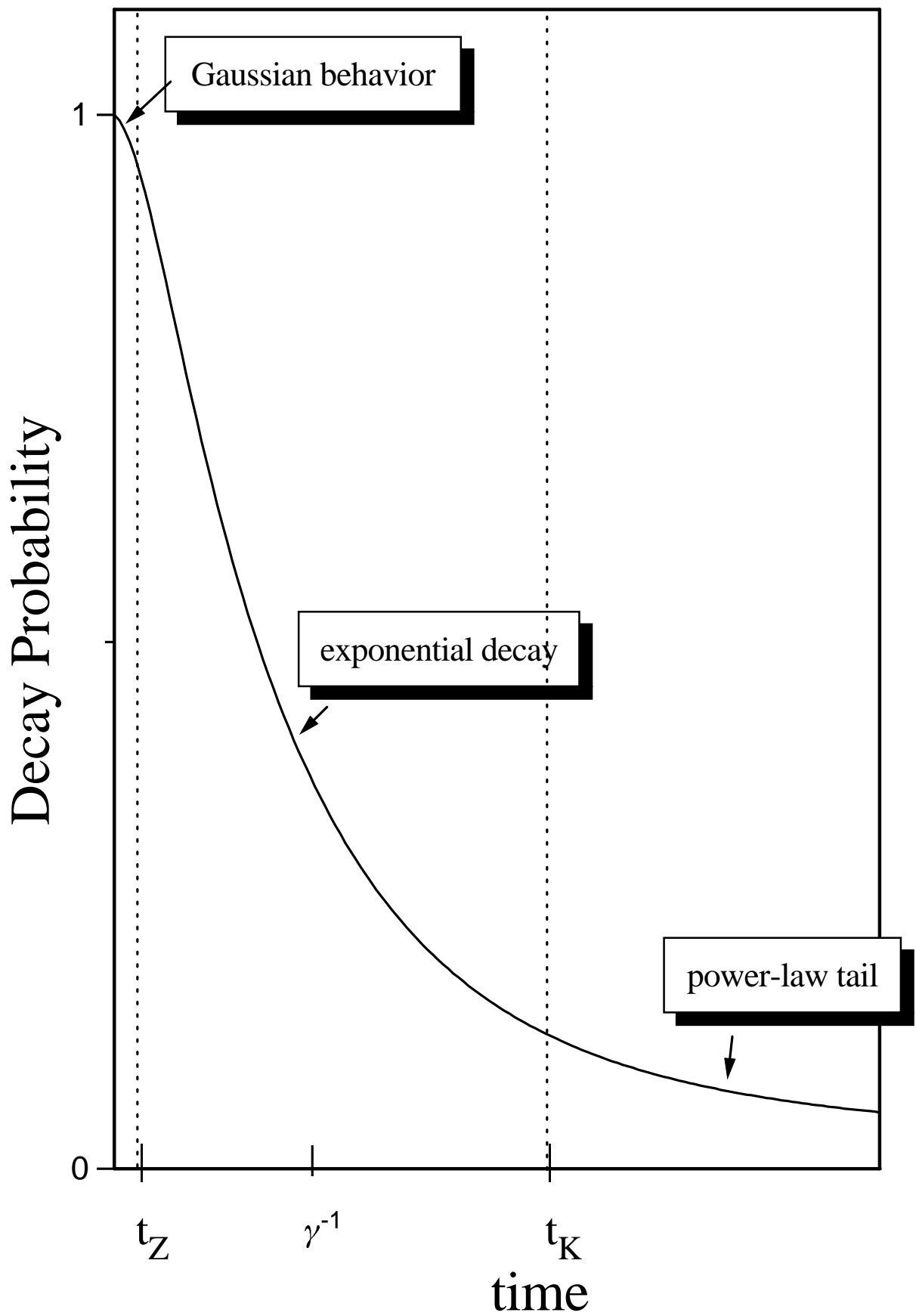
\end{document}